# Self-seeded photon acceleration by electron beam-driven transition radiation

Chaolu Ding,[1,2] Xuesong Geng,[1,*] and Liangliang Ji [1,**]

[1] *State Key Laboratory of Ultra-intense laser Science and Technology,*
*Shanghai Institute of Optics and Fine Mechanics (SIOM),*
*Chinese Academy of Sciences (CAS), Shanghai, 201800, China.*
[2] *Center of Materials Science and Optoelectronics Engineering,*
*University of Chinese Academy of Sciences, Beijing, 100049, China.*

*xsgeng@siom.ac.cn
**jill@siom.ac.cn

**Abstract:**

Photon acceleration (PA) driven by ultra-relativistic electron beams offers a promising approach to generating high-power, high-frequency coherent radiation sources. While current methods typically rely on external optical laser pulses injected into beam-driven plasma wakefields, they face significant challenges in synchronization and alignment between electron accelerators and laser systems. We propose utilizing transition radiation (TR) generated by the drive electron bunch transversing the vacuum-gas interface as the seed photons of PA. Using a 1 GeV electron bunch, we demonstrate acceleration of TR from 4.4 μm to 184 nm in 1.6 mm of two-stage uniform plasma, achieving more than a 20-fold frequency boost. Further frequency increases can be achieved with optimized setups. This scheme addresses the synchronization and alignment issues present in previous approaches, providing a practical path toward beam-driven photon acceleration.

## 1. Introduction

Photon acceleration (PA), which exploits frequency up-shift of light co-propagating with a decreasing refractive index, offers a promising approach for generating high-power, short-wavelength radiation [1,2]. In this process, photons gain energy from plasma waves while the pulse undergoes temporal compression [3], potentially enabling significant power enhancement. Dissipation of the driver energy and the limited acceleration length due to photon dephasing restrict frequency up-shifts to within 25% in optical regime experiments [4–11]. To mitigate such effects, flying focus laser [12,13] and tailored plasma density profiles [14–18] have been proposed theoretically, showing that a 50 GeV electron beam can upshift the frequency of a Gaussian or vortex pulse from 800 nm to 36 nm. However, two primary challenges persist: the complex synchronization and alignment between conventional electron accelerators and laser systems, and the high energy required for the driver.

To overcome these challenges, we propose the self-seeded photon acceleration (SPA) concept that eliminates the need for an external seed laser. The SPA utilizes transition radiation (TR) generated by an electron bunch passing through a vacuum-gas interface as the seed photon, inherently ensuring perfect synchronization with the plasma wake. Using quasi-3D Particle-In-Cell (PIC) simulations, we demonstrate that radially polarized TR can be boosted from 4.4 $\mu m$ (mid-infrared, mid-IR) to 184 nm (ultraviolet, UV) in a plasma wake driven by a 1 GeV electron beam with parameters expected to be achievable at facilities like FACET-II [19,20]. The wavelength can be further boosted with optimized designs. This simple configuration paves the path towards the first proof-of-concept demonstration of significant photon acceleration from infrared to ultraviolet wavelengths.

This article is structured as follows: Section 2 introduces the SPA scheme and fundamental concepts of TR. Section 3 describes simulation setup. Section 4 presents PIC simulation results, and Section 5 discusses the scheme's optimization and potential applications.

## 2. Self-seeded Photon Accelerating Scheme

The interaction scheme of SPA is illustrated in Figure 1(a), using a 1 GeV electron beam with a 5 nC beam charge. The driven beam takes a pancake shape, characterized by a transverse size significantly larger than its longitudinal size ($\sigma_z = 0.5$μm, $\sigma_r = 15$μm). Such electron sources are achievable at the forthcoming FACET-II facility [19,20]. The plasma converter consists of a photon generator (linear up-ramp), a photon accelerator (plateau), and a down-ramp. Plasma density and length scale are dependent on the beam density. In our case, we set both beam and plasma density to $n_b = n_p = 1.74 \times 10^{19} cm^{-3}$. As shown in the following, the width-to-length ratio of the drive beam $\sigma_r/\sigma_z \gg 1$ situates our scheme within the coherent transition radiation (CTR) regime, where the diffraction length is large $\sigma_r^2/\sigma_z \gg \sigma_z$ [21]. The beam-to-plasma density ratio $n_b/n_p = 1$ corresponds to the so-called relativistic transition radiation (RTR) [22], where electrons in plasma undergo relativistic oscillations, resulting in the formation of high-density electron sheets and intense radiation pulse trains. These requirements are essential in producing seed photons.

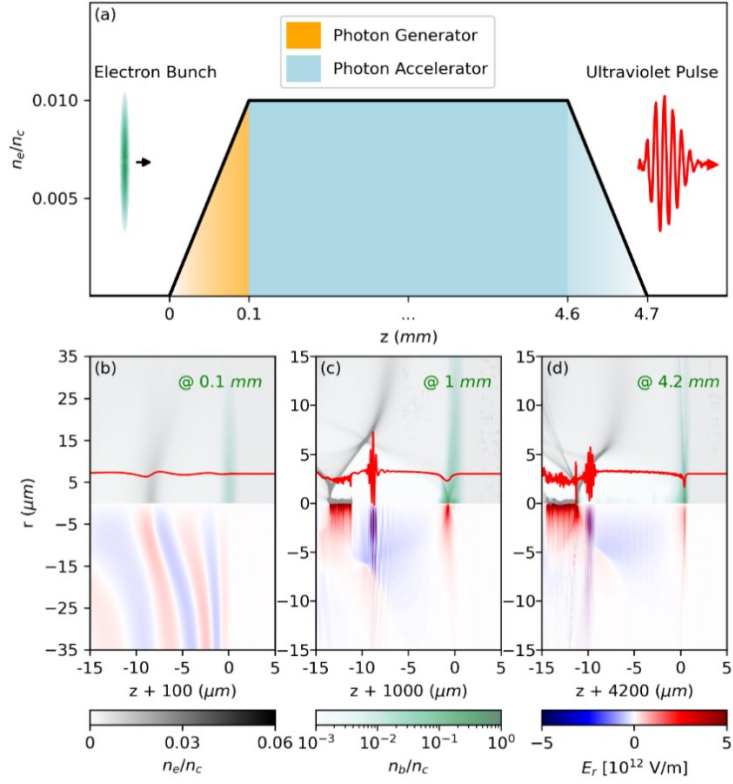

Fig. 1. Schematic diagram of the beam–plasma interaction configuration. (a) Ultraviolet pulse generation occurs when an ultra-short electron beam, with a density equal to the plasma density $n_b = n_p$, impinges on a uniform plasma with up- and down-ramp of 0.1 mm. (b-d) The evolution of electron beam density (green), plasma electron density (gray), and radial electric field in $\theta = 0$ plane at propagation distances of 0.1 mm, 1 mm, and 4.2 mm, respectively, with extracted electric field profiles at $r = 3\mu m$ shown in red (a.u.).

As the beam enters the up-ramp plasma-the "photon generator", the forward-propagating CTR is generated in the mid-infrared spectrum, induced by strong plasma electrostatic field drive thin electron layers at the vacuum-plasma interface, which will be shown later. This TR serves as the seed pulse for subsequent photon acceleration, as illustrated in Figure 1(b). In the plateau region, the quasi-planar beam drives a quasi-one-dimensional plasma wake, forming a refractive index gradient structure. Photons sitting in the co-moving index gradient experience frequency upshift as described by [10]

$$\frac{\Delta\omega}{\omega_0} \approx -\frac{\omega_p^2}{2\omega_0^2}\frac{c}{n_0}\int_{-\infty}^{\infty}\frac{\partial n}{\partial \zeta}dt, \quad (1)$$

where $\Delta\omega/\omega_0$ represents the fractional change in frequency relative to its initial central frequency $\omega_0$ and $\omega_p = \sqrt{\frac{e^2 n_0}{m_e \epsilon_0}}$ denotes the plasma frequency with the elementary charge $e$, the electron rest mass $m_e$, and the vacuum permittivity $\epsilon_0$, respectively. Here, $n$ and $n_0$ are the perturbed and initial electron densities. The variable $\zeta = z - ct$ is the relative phase between the pulse and the driver, where $z$ and $t$ represent the spatial position and time in the lab frame, and $c$ is the speed of light in a vacuum. When the radiation frequency $\omega_0$ increases, maintaining the acceleration gradient requires a higher density gradient $\frac{\partial n}{\partial \zeta}$.

Multi-dimensional effects come into play in later stages, where plasma acts as a plasma lens [23,24] that focuses the drive beam. As depicted in Figure 1(c), when the equivalent beam density exceeds the background plasma density, the wakefield enters the blowout regime [25]. Figure 1(d) illustrates that the bubble sheath associated with steep density spike provides a higher density gradient, i.e., a refractive index gradient, thereby further boost the frequency up-shift of the self-generated seed radiation.

## 3. Simulation Setup

To model the SPA scheme, we perform spectral, quasi-three-dimensional PIC simulations using the Fourier-Bessel Particle-In-Cell code (FBPIC) [26]. It solves Maxwell's equations using a spectral approach in cylindrical coordinates, mitigating spurious numerical dispersion. It decomposes the electromagnetic field into a set of 2D radial grids using azimuthal Fourier decomposition, where the angular component is expressed through the azimuthal Fourier expansion of $e^{im\theta}$. The moving simulation box size is (40 μm, 75 μm) with cell numbers of (4000, 500) in the (z, r) directions. Given the system's cylindrical symmetry, two azimuthal modes ($N_m = 2$) suffice to capture the primary physics and the non-axisymmetric problems. We allocate (2, 2, 8) particles per cell (PPC) in the $(z, r, \theta)$ directions for each plasma species. This configuration corresponds to $6.40 \times 10^7$ macroparticles per species, while the electron bunch comprises $1.92 \times 10^7$ macroparticles. The plasma density is constant at $n_p = 1.74 \times 10^{19} cm^{-3}$ over a 4.5 mm plateau, with linear ramps of 0.1 mm at both the rising and falling edges.

## 4. Results

### 4.1 Acceleration of transition radiation

The evolution of the radiation spectrum is shown in Figure 2(a), calculated by Fourier transform of the $E_r$ field along the z-direction and summed over radius. The electrostatic component of the wakefield is eliminated after filtering. The wavelength decreases rapidly from 4.4 μm to 267 nm, stabilizes, and then declines slightly to 201 nm. We categorize the photon acceleration process into three stages: I) the quasi-1D plasma wake stage (0.1 mm to 0.8 mm); II) the blowout-regime stage (0.8 mm to 3.5 mm); and III) the rephasing stage (3.5 mm to 4.7 mm). The existence of stage II indicates the potential of accelerating to shorter wavelengths with dedicated setups.

As the electron energy decreases, the pulse energy also drops rapidly during the first stage and then remains almost constant, as illustrated in Figure 2(b). We calculate the central energy of the radiated photons and obtain the photon number from the total pulse energy. From Figure 2(c), one notices that while the average photon energy is boosted rapidly, the total photon number declines due to dephasing. The drop in pulse energy is primarily induced by the limited number of trapped photons due to dephasing in the uniform plasma. Figure 2(d) illustrates the evolution of peak intensity and the normalized amplitude $a_0 = \frac{eE}{m_e c \omega_0}$. The pulse intensity fluctuation stems from inefficient photon trapping, light diffraction, and self-focusing effects within the evolving plasma wake. As the pulse exits the plasma down-ramp, it attenuates, decreasing in peak intensity from $6.58 \times 10^{17}$ to $2.72 \times 10^{17}$ $W/cm^2$ due to diffraction, corresponding to normalized field strength of $a_0 \approx 0.07$. The energy conversion efficiency, defined as the ratio of pulse energy (1.58 mJ at 4.7 mm) to driver energy loss, reaches 0.057%.

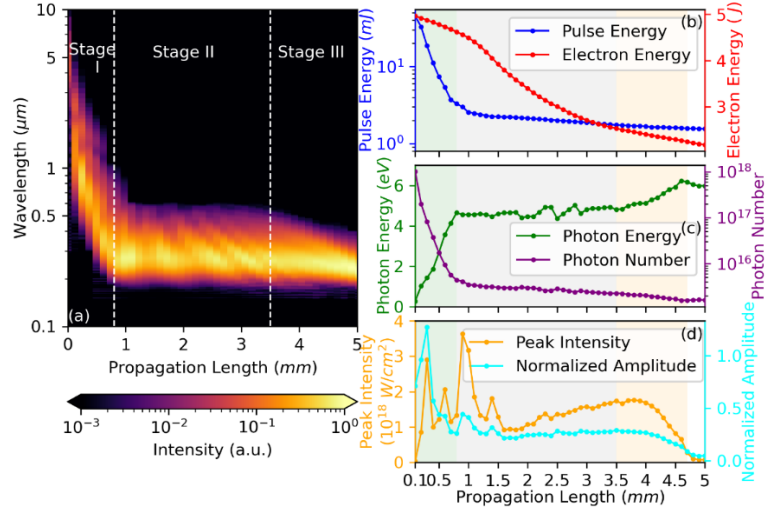

Fig. 2. Photon acceleration process along the propagation length of the electron beam in plasma. (a) The evolution of the wavelength spectrum as a function of the electron beam propagation distance in plasma, divided into three stages at 0.8 mm and 3.5 mm. (b) The filtered light pulse energy (blue line) and the total electron beam energy (red line). (c) The single-photon energy (green line) and the total photon number (purple line). (d) The light pulse peak intensity (orange line) and the dimensionless amplitude a0 (cyan line). Stage I/II/III are shaded by green, gray and orange.

The detailed evolution of the accelerated wave packet is analyzed in Figure 3, revealing distinct chirp characteristics as shown in Figure 3(a-c). Figure 3(a) shows that the initial seeded radiation pulse exhibits no chirp. Figure 3(b) presents a positive chirp, arising from a higher accelerating gradient in the pulse tail compared to the leading front during the first stage. Figure 3(c) illustrates a chirp reversal during the third acceleration stage, where larger-radius pulse components undergo initial frequency up-conversion before refracting inward, resulting in concentrated high-frequency components at the leading edge. In other words, the high-frequency components come from the bubble edge far from center.

To further explain the above results, the relative position of the wake and the TR is shown in Figure 3(d-f). In Figure 3(d), the witness pulse experiences a moderate density gradient within the quasi-one-dimensional plasma wake and a relatively large accelerating gradient due to the small central frequency $\omega_0$, as the relative frequency shift is inversely proportional to the square of $\omega_0^2$ according to Equation (1). The initial wavelength at the beginning of the first stage of photon acceleration is 4 $\mu m$ at $r = 3\mu m$. After propagation of 0.9mm, as shown in Figure 3(e), the electrons are entirely expelled into the blowout regime due to the self-focusing of the driving bunch, resulting in no change in the pulse wavelength from 1mm to 3.5mm of propagation, indicating significant potential of being further accelerated. At 4.2mm, as shown in in Figure 3(f), the driver depletes energy during the second stage, causing the bubble to contract about 2.1 $\mu m$ in the third stage, trapping the pulse in a steep plasma density gradient again, enabling photons to accelerate efficiently to shorter wavelength compared to the first stage. The average wavelength along $r$ further decreases from 258 nm to 201 nm in the third stage. We emphasize the importance of the third stage, where the bubble edge can also be utilized for photon acceleration, resulting in a significant acceleration gradient. Figure 3(g) demonstrates that the intensity increases as the wavelength decreases during the photon acceleration process.

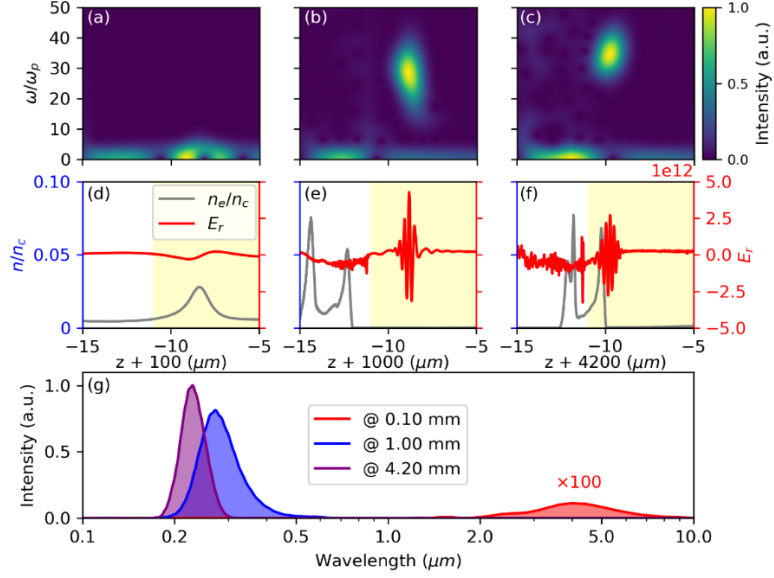

Fig. 3. Extracted radial electric field and plasma density distributions at $r = 3\mu m$ for electron beam positions indicated in Figures 1(b-d). (a-c) The time-frequency analysis of $E_r$ obtained via short-time Fourier transforms. (d-f) The evolution of radial electric field (red line) and plasma density (gray line) profiles at corresponding positions. (g) The wavelength spectra of $E_r$ within the shaded yellow regions in panels (d-f).

### 4.2 Generation of coherent transition radiation

For an ultrarelativistic electron beam with a Gaussian density distribution $n_b = n_0 \cdot \exp\left(-\frac{r^2}{2\sigma_r^2} - \frac{z^2}{2\sigma_z^2}\right)$, the self-generated field surrounding the beam primarily comprises the radial electric field $E_r$ and the azimuthal magnetic field $B_\theta$, which can be approximated as

$$E_r(r,z) \approx cB_\theta \approx -\frac{n_0 e}{\epsilon_0}\frac{\sigma_r^2}{r}\left(1 - e^{-\frac{r^2}{2\sigma_r^2}}\right) e^{-\frac{z^2}{2\sigma_z^2}} \qquad (2)$$

When such an electron beam traverses the vacuum-plasma interface, the electrons are first pushed outward radially by the radial electric field, acquiring an initial radial velocity $v_r$. Subsequently, the $v_r \times B$ force turns the electrons forward, leading to the compression of tenuous electrons at the front of the up-ramp plasma at $t = 50.0$ fs, as illustrated in Figure 4(a). Here, $t = 0$ corresponds to the moment when the center of the electron beam reaches the vacuum-plasma interface at $z = 0$. Meanwhile, a quasi-static field builds up due to charge separation, as shown in Figure 4(b). It grows continuously such that the electrostatic force excreted on electrons surpasses the ponderomotive force. The electron layer is thus retracted backwards, as shown in Figure 4(c).

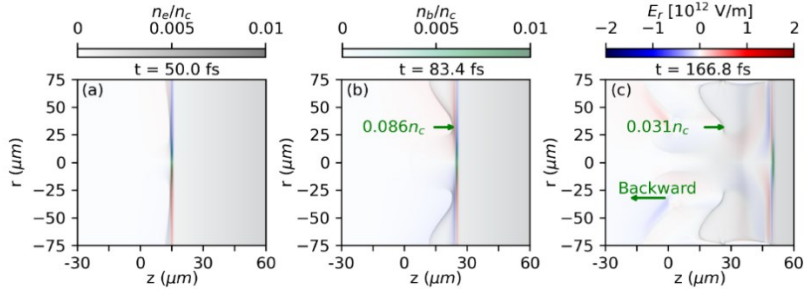

Fig. 4. Spatiotemporal evolution of electron density and radial electric field. (a-c) The electron beam density (green), plasma electron density (gray), and radial electric field distributions at 50.0 fs, 83.4 fs, and 166.8 fs, respectively.

For concrete explanation of the effect, the radiation field of the electron in the electron layer is shown in in Fig. 5, calculated by evaluating the Liénard–Wiechert potential [27] from the electron trajectories with the open source package PyCharge [28]. According to classical radiation theory, the radiation field of a moving electric point charge is described by

$$\vec{E}(\vec{r},t) = \frac{q}{4\pi\epsilon_0} \cdot \left\{ \frac{|\vec{n}|}{(\vec{n}\cdot\vec{u})^3} \cdot [\vec{n} \times (\vec{u} \times \vec{a})] \right\}_{ret}. \quad (3)$$

Here, $\{...\}_{ret}$ denotes the retarded time, $\vec{n} = \vec{r} - \vec{r}_{ret}$ is the vector from the particle's position at the retarded time $\vec{r}_{ret}$ to the observation point $\vec{r}$, $\vec{u}$ is defined as $\vec{u} = c\frac{\vec{n}}{|\vec{n}|} - \vec{v}$, and $\vec{v}$ and $\vec{a}$ represent the particle's velocity and acceleration at the retarded time, respectively. For an electron in the $x = 0$ plane ($\theta = 0$ in cylindrical coordinates), the radial radiation field simplifies to

$$\vec{E}_y(\vec{r},t) = \frac{q}{4\pi\epsilon_0} \cdot \left\{ \frac{|\vec{n}|}{(\vec{n}\cdot\vec{u})^3} \cdot n_z(a_z u_y - u_z a_y) \right\}_{ret}. \quad (4)$$

This formula indicates that both longitudinal and radial motions contribute to the radial radiation field. Figure 5(a) shows the calculated radiation field for a representative electron initiated at $z = 10$μm and $r = 25$μm, with its trajectory marked by a green line. The radiation field pattern originates from two key positions: the initial point and turning point. The electron's initial acceleration generates a positive half-cycle in the leading wavefront, while its deceleration at the turning point produces a negative half-cycle. These two half-cycles constitute one complete cycle of the foremost radiation wavefront. In Figure 5(b), the electron first acquires radial velocity and subsequently gains longitudinal velocity when the normalized radial velocity $\beta_y = v_y/c$ reaches 0.1. Between 40 and 80 fs, the electron propagates in the z direction at a nearly uniform velocity of $0.87c$. After this period, the electron lags behind the beam by $1.56\mu m$, detaching from its self-field. Subsequently, the electron undergoes deceleration in both longitudinal and radial directions, as shown in Figure 5(c). The strong longitudinal and mild radial decelerations arise from the characteristic quasi-static field induced by the annular electron layer.

The total radiation field is calculated using test electrons, which are uniformly sampled along the z- and r-directions at the $x = 0$ plane with intervals of $\Delta z = 0.1$μm and $\Delta r = 0.5$μm. Figure 5(d) compares the normalized radiation fields obtained from calculations (upper half-panel) and the normalized PIC results (lower half-panel). The radiation wavefront replicates the electron layer's shape at emission, in accordance with the Huygens-Fresnel principle. Electrons in the initial z-plane emit spherical waves forming a forward-propagating plane wavefront, while the retracting annular electron layer generates tilted-pulse radiation propagating backward along the direction shown by the green arrow. Figure 5(e) displays the normalized radiation field and total field at $r = 40$μm using orange and blue lines, respectively. The leading positive half-period radiation peak coincides with the electron beam's maximum density, with the radial radiation field opposing the beam self-field. This behavior supports the transition radiation mechanism, wherein the radiation field induced by plasma electron motion counteracts the beam self-field. Both curves exhibit consistent behavior in vacuum ($z < 0\mu m$), while their difference in the plasma region ($z > 0\mu m$) stems from the PIC results incorporating both radiation and Coulomb fields, compared to only the test electron radiation field. Figure 5(f) illustrates the extracted plasma and beam density distributions. The negative half-period radiation, having separated from the electron beam, propagates with the decreasing plasma density gradient over time. This configuration results in photon deceleration, extending the wavelength into the acceleration phase.

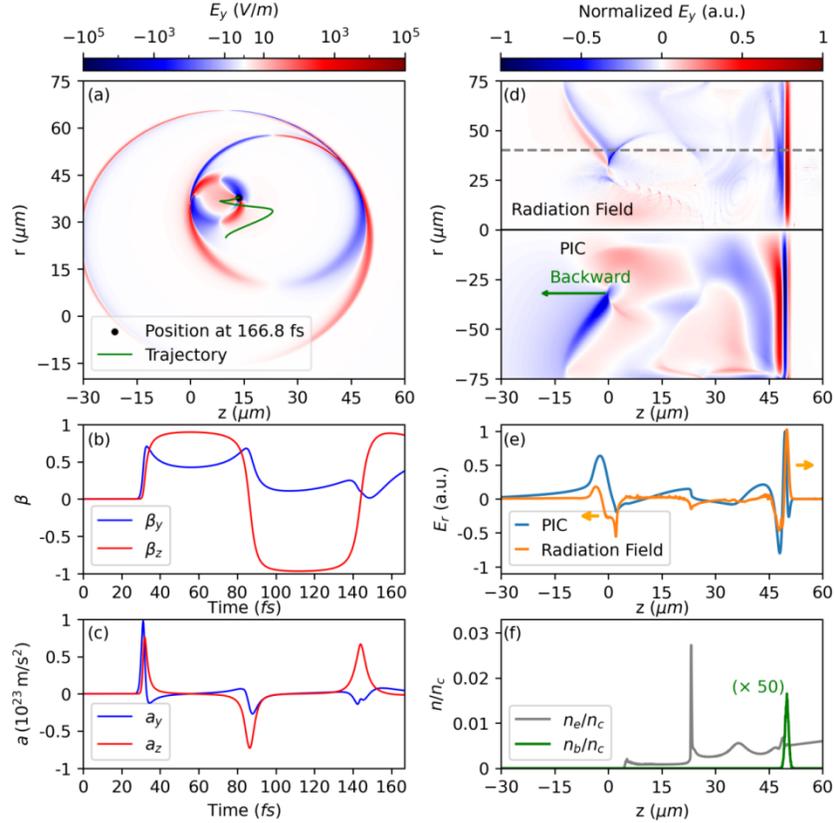

Fig. 5. Radiation field calculated using the Liénard–Wiechert potentials for single and all test electrons at t = 166.8 fs. (a-c) depict a typical test electron. (a) The radial radiation field distribution with electron position (black dot) and trajectory (green line). (b) The temporal evolution of normalized radial velocity $\beta_y = v_y/c$ and longitudinal velocity $\beta_z = v_z/c$. (c) The temporal profiles of radial acceleration $a_y$ and longitudinal acceleration $a_z$. (d–f) present the results for all test electrons. (d) Comparison of normalized radiation field (upper half-plane) and normalized total field $E_y$ from PIC simulations (lower half-plane), with field extraction line at $r = 40\mu m$ (gray dashed). (e) The normalized radiation field (orange line) and total field (blue line) profiles, with radiation field propagation direction indicated by orange arrows. (f) The beam density (green line, multiplied by 50) and plasma density (gray line) distributions at $r = 40\mu m$.

## 5. Discussion

The second stage of the uniform plasma case leads to reduced energy transfer efficiency. In the blowout regime, the pulse resides inside the bubble, thereby inhibiting an increase in pulse frequency. Consequently, both the pulse and the electron beam lose nearly half of their initial energy during the second stage. To improve energy transfer efficiency, we try a two-stage photon acceleration scheme using a customized ascending density ladder plasma profile. As illustrated in Figure 6(a), the plasma exhibits a linear up-ramp and down-ramp of 0.1 mm, along with two plateaus, each measuring 0.7 mm longitudinally. The plasma density of the first plateau is 0.01 $n_c$, and the second plateau has a density three times greater than the first. After pulse dephasing in the first wake, the plasma density triples, causing the pulse to transition to the second wake, where additional frequency up-shifting occurs at the plasma wake's edge. As shown in Figure 6(c), when the electron beam enters the second plasma density plateau at 0.85 mm, the pulse slips from the first to the second plasma wake. Figure 6(d) displays the radial electric field and density distributions at $r = 5\mu m$, corresponding to Figure 6(c). It is observed that the wedge of the plasma wake creates a decreasing plasma density gradient, and the pulse is precisely positioned there, leading to the rephasing of photons within the second plasma wake. Unlike Figure 4(a), Figure 6(b) demonstrates that the wavelength spectrum continues to decrease during the second stage instead of forming a plateau. Upon exiting the plasma, the pulse energy and wavelength are 1.69 mJ and 184 nm, respectively, yielding an energy transfer efficiency of 0.15%, higher than the uniform case. In practical applications, the density transition between Stage 1 and Stage 2 occurs gradually rather than abruptly; therefore, a $100\mu m$ density ramp is introduced between the two stages. This modification slightly reduces performance, producing a 209 nm, 1.4 mJ pulse with an efficiency of 0.116%.

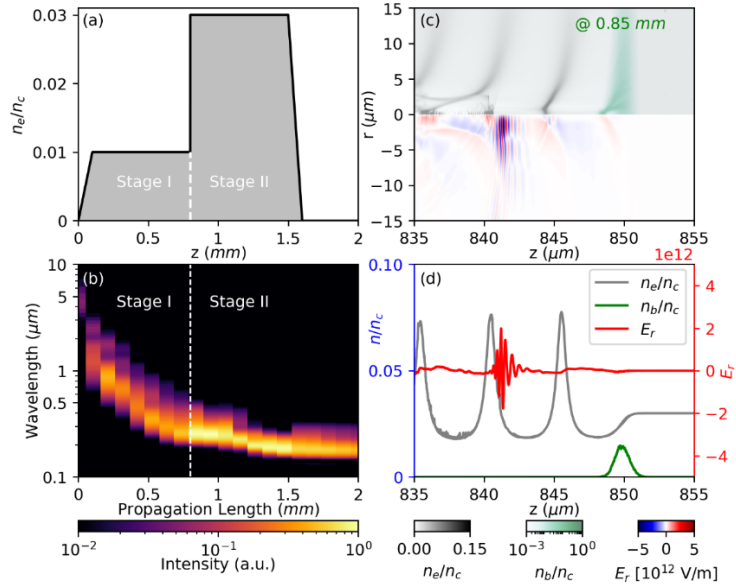

Fig. 6. Photon acceleration in a tailored plasma. (a) The longitudinal plasma density profile. (b) The evolution of the radiation spectrum as a function of the electron beam propagation distance in plasma, divided into two stages at 0.8 mm. Stage I/II are distinguished by the white dashed line. (c) The electron beam density (green), plasma electron density (gray), and radial electric field

in the $\theta = 0$ plane at 0.85 mm propagation distance. (d) The extracted radial electric field and density profiles at $r = 5\mu m$ corresponding to beam positions shown in panel (c).

The pulse wavelength can be flexibly tuned by adjusting the plasma length. When an electron beam interacts with a short plasma, it generates a long-wavelength infrared vector pulse due to the inherently radially polarized nature of TR. This development is promising in various potential applications [29,30] of vector infrared pulses. On the other hand, the upper limit of frequency shift is determined by the gap between plasma electrons [31] (i.e., $\lambda \gg n_0^{-\frac{1}{3}}$). For plasma density of $10^{19}$ cm$^{-3}$, which can be reached at the end of the bubble in the blowout regime, the minimum attainable wavelength is approximately 50 nm

## 6. Conclusion

In summary, we present a proof-of-concept demonstration of the interaction between an ultra-short relativistic electron beam and a tenuous plasma with a density ratio of approximately $n_b/n_p = 1$ using FBPIC. We found that the forward CTR generated at the front vacuum-plasma interface can be guided within the plasma wake, and the radiation frequency is modulated from IR to UV. Furthermore, we enhance energy transfer efficiency by tailoring plasma density profile. This study is the first to integrate the concepts of PA and TR within an ultra-short electron beam parameter regime and to develop the SPA scheme. It provides a wavelength-tunable vector radiation source, with its minimum wavelength constrained by the electron beam energy and plasma density. By incorporating advanced plasma engineering techniques, this scheme holds significant potential for generating radially polarized UVs below 100 nm. This research is of significant importance in fundamental physics and has substantial implications for fields such as plasma optics and astrophysics.

**Funding:** This work is supported by the National Natural Science Foundation of China (Nos. 12388102, 12374298 and 12304384), the Strategic Priority Research Program of Chinese Academy of Sciences (No. XDB890303), the National Key R&D Program of China (No. 2022YFE0204800), Shanghai Science and Technology Development Foundation (Grant No. 22YF1455100).

**Acknowledgment:** The authors appreciate the fruitful discussions with Professors Alexander Pukhov and Hao Peng.